\documentclass[aps,prl,twocolumn,showpacs]{revtex4}
\usepackage{epsfig}
\usepackage{graphicx}

\begin{document}

\DeclareGraphicsExtensions{.eps,.EPS}

\title{Probing spin dynamics from the Mott insulating to the superfluid regime in a dipolar lattice gas}
\author{A. de Paz$^{1,2}$, P. Pedri$^{1,2}$, A. Sharma$^{1,3}$, M. Efremov$^{1,4}$, B. Naylor$^{2,1}$, O. Gorceix$^{1,2}$, E. Mar\'echal$^{2,1}$, L. Vernac$^{1,2}$, B. Laburthe-Tolra$^{2,1}$}
\affiliation{1 Universit\'e Paris 13, Sorbonne Paris Cit\'e, Laboratoire de Physique des Lasers, F-93430
Villetaneuse, France; 2 CNRS, UMR 7538, LPL, F-93430 Villetaneuse, France; 3 Department of Physics and
Astronomy, University of Sussex, Brighton BN1 9QH, UK; 4 Institut fur Quantenphysik
Universitat Ulm Albert-Einstein-Alle 11 89081 Ulm, Germany}

\begin{abstract}
We analyze the spin dynamics of an out-of-equilibrium large spin dipolar atomic Bose gas in an optical lattice. We observe a smooth crossover from a complex oscillatory behavior to an exponential behavior throughout the Mott to superfluid transition. While both of these regimes are well described by our theoretical models, we provide data in the intermediate regime where dipolar interactions, contact interactions, and super-exchange mechanisms compete. In this strongly correlated regime, spin dynamics and transport are coupled, which challenges theoretical models for quantum magnetism.

\end{abstract}
\date{\today}
\maketitle

Dipolar atoms and molecules loaded in optical lattices are a promising platform to study quantum many-body physics \cite{reviews,ferlaino2015}, and in particular  quantum magnetism \cite{dipolequantummag1,dipolequantummag2,dipolequantummag3,dipolequantummag4,dipolequantummag5,dipolequantummag7}. In dipolar systems direct spin-spin interactions are provided by dipole-dipole interaction (DDI) without relying on
super-exchange mechanism \cite{auerbach}. Although magnetization changing collisions associated to the anisotropic character of dipolar interactions may introduce interesting exotic quantum phases \cite{Buchler,Carr,Zoller,Syzranov}, these off-resonant processes are often negligible. Then dipolar interactions reduce to the following Hamiltonian:
\begin{equation}
V_{\rm sc}=\frac{d^2}{r^3} \left(S_1^zS_2^z  -
\frac{S_1^+ S_2^- + S_1^- S_2^+}{4}
\right)(1-3\cos^2(\theta_{1,2}))
\label{heisenberglike}
\end{equation}
where $d^2 = \mu_0 / 4\pi (g \mu_B)^2$ ($\mu_0$ being the magnetic permeability of vacuum, $g$ the Lande factor, $\mu_B$ the Bohr magneton), $r$ is the distance between atoms, $\theta_{1,2}$ the angle between the magnetic field and the interatomic axis, and $S_i^{\pm,z}$ are the spin operators acting on atom $i$.
This Hamiltonian, known as the secular dipolar Hamiltonian in the context of Nuclear Magnetic Resonance \cite{abragam}, bears strong similarities with the XXZ model of quantum magnetism \cite{auerbach}. 

Experimental investigation of such spin Hamiltonians has recently started, with dipolar molecules \cite{exchangeye}, magnetic \cite{3DSpinExchange} and Rydberg \cite{barredo2015} atoms, which is triggering important interest \cite{dipolequantummag7,Buchler,Carr,Zoller,Syzranov,interest}. While these studies have focused in a localized regime where the particles are pinned to a well-defined position, in this paper we investigate the case where magnetic atoms are free to move in an optical lattice. Thus spin dynamics and transport are coupled and conform to an intriguing interplay between super exchange mechanisms and dipolar spin exchange. Our experiment provides first data in this regime which challenges theoretical description. 

\begin{figure*}
\centering
\includegraphics[width= 17.2 cm]{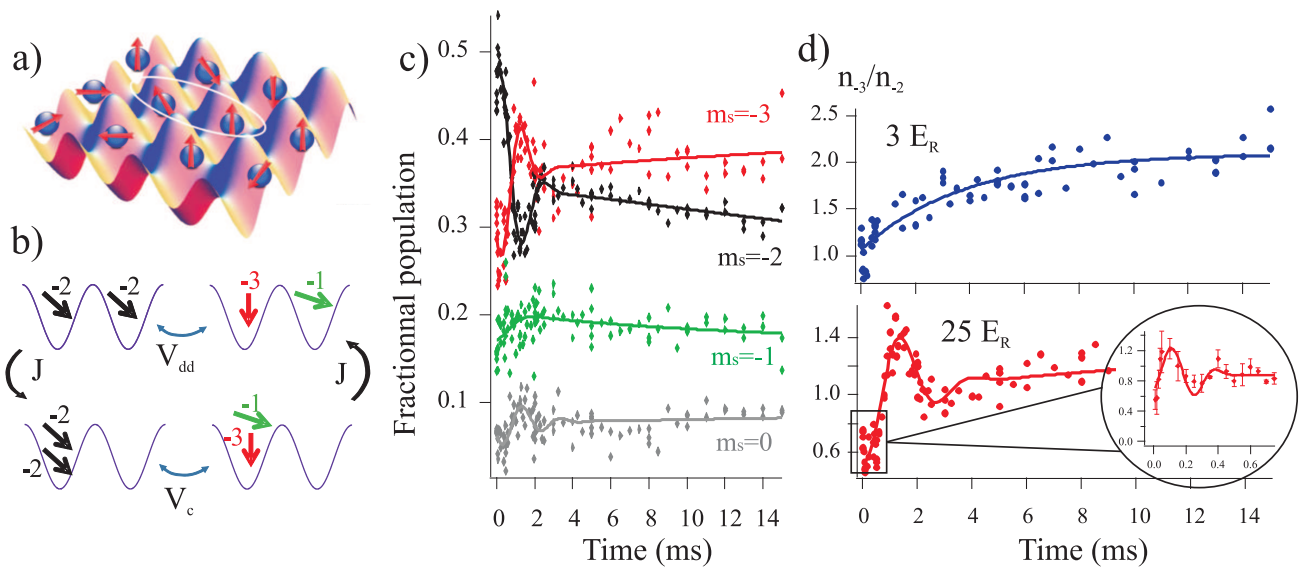}
\caption{a) Simple representation of the system in the Mott phase with one atom per site, with atoms interacting via DDI (white ellipse); b) sketch illustrating competition between exchange due to DDI ($V_{dd}$) and tunneling ($J$) assisted spin-exchange due to contact interactions ($V_c$); c) typical measurement of the spin components as a function of time; d)
time evolution of observable $n_{-3}/n_{-2}$ for two extreme regimes (top : superfluid, bottom : Mott). Lines are guides for the eye resulting from fits.}
\label{principe_data2}
\end{figure*}

We study spin exchange dynamics of magnetic chromium $^{52}$Cr bosonic atoms loaded in a 3D
optical lattice, across the Mott to superfluid transition \cite{greiner2002}.
We observe, as a function of the lattice depth, a crossover between two distinct behaviors. In the Mott phase, spin dynamics displays a complex oscillatory behavior. We identify two distinct frequencies, one associated to on-site spin-dependent contact interactions and the other to inter-site DDI. In the superfluid regime, spin dynamics shows an exponential behavior which results from an interplay between contact and dipolar interactions. The regime at intermediate lattice depth is particularly interesting because super-exchange mechanisms also contribute to the spin dynamics. For example an atom may tunnel into an already occupied site and interact with another atom by spin dependent contact interaction; this can trigger spin changing collisions (see Fig.\ref{principe_data2}b)). 
It is extremely challenging to simulate the many-body quantum spin dynamics in this intermediate regime where three exchange mechanisms compete (associated to DDI, contact interactions, and super-exchange). We experimentally find that the oscillations observed in the Mott phase survive with a reduced amplitude.

We perform our experiment with a spin-3 $^{52}$Cr Bose-Einstein condensate,
comprising 10$^4$ atoms loaded into an anisotropic optical lattice \cite{3DSpinExchange}. As the lattice depth is spanned, we observe the superfluid to Mott transition, at a typical lattice depth of 12 $E_R$ (where $E_R$ is the recoil energy). For our experimental parameters, the system in the Mott phase consists of a core with two atoms per site, surrounded by a shell with one atom per site.

To initiate spin-dynamics atoms are transfered into the first single-particle excited Zeeman state $m_s=-2$ using the tensor light-shift of a 427.85 nm light pulse \cite{3DSpinExchange}. We then measure, after a variable hold time $t$, the spin populations by means of a Stern-Gerlach procedure. A typical result is plotted in Fig.\ref{principe_data2}c).
Only the  $m_s=-3$, $-2$, $-1$ and $0$ spin components are significantly populated as the system evolves. The populations display a rather complex behavior as a function of time. In order to simplify the discussion, we focus our attention onto the observable given by the ratio $n_{-3}/n_{-2}$ of $m_s=-3$ and $m_s=-2$ populations, since they are the most populated components.

\begin{figure}
\centering
\includegraphics[width= 7 cm]{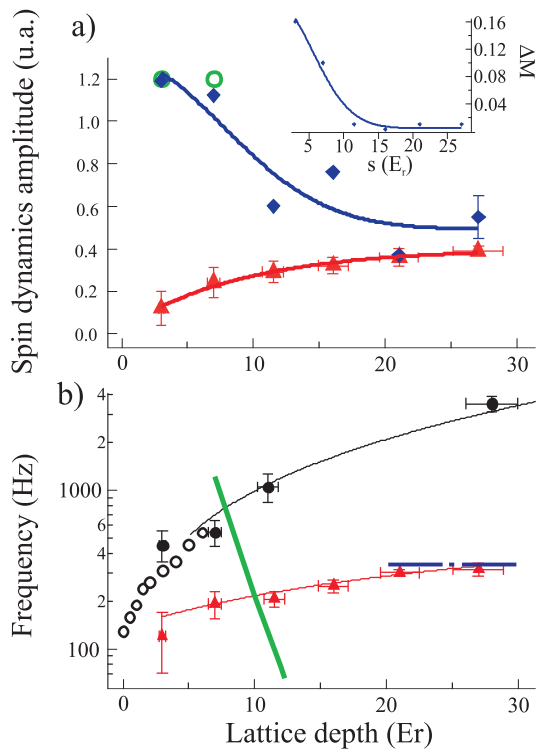}
\caption{Spin dynamics amplitudes and frequencies as a function of the lattice depth. a) Amplitude of the exponential dynamics (blue diamonds) and slow oscillation (red triangles). Green circles are results of numerical simulations. Inset: variation of the magnetization over 20 ms. Solid lines are guides to the eye.
b) Frequency of fast (black points) and slow (red triangles) oscillations. The black top solid line corresponds to spin-exchange frequency associated to intrasite contact interactions, while the black open circles correspond to a numerical simulation of Gross-Pitaevskii equation. The bottom red curve is a guide to the eye. The blue dot-dashed line shows the prediction in the Mott regime (see text).  The frequency of super-exchange process is given by the green solid line. Error bars in frequency and amplitude result from the statistical uncertainty in the fits.}
\label{AmpFreq}
\end{figure}

We have plotted in Fig.\ref{principe_data2} d) the typical results corresponding to the extreme regimes of high and shallow lattice depths, showing quite different spin dynamics. In the Mott phase we observe at short times ($<0.5$ms, see inset) a strongly damped oscillation, and then at longer times a second oscillation. In the superfluid phase the spin dynamics is better described by an exponential. All these features are present in the data  from 3 $E_R$ to 25 $E_R$. We plot in Fig.\ref{AmpFreq}a the amplitudes of the exponential and of the slow oscillation, and in Fig.\ref{AmpFreq}b the frequencies of the fast and slow oscillations. When reducing the lattice depth, we observe that both oscillation frequencies decrease and they become closer to each other. Oscillations do survive at lattice depth slightly below the Mott to superfluid transition. However, for very shallow lattice depth, the oscillations at low frequency are barely visible and the spin dynamics is mostly exponential. 

For most of the data shown in Fig. \ref{AmpFreq} magnetization is constant (see the inset of Fig.\ref{AmpFreq}a)). The stability of the magnetization for large lattice depths indicates that dipolar relaxation is suppressed, which arises because the energy released in a dipolar relaxation event (the Larmor energy $\approx 30$ kHz) does not match band excitation ($>50 $kHz) \cite{Resonances}. At small lattice depth, dipolar relaxation is not completely suppressed (as the first excited band has an energy close to the Larmor energy). However, this effect remains rather small, and is neglected in the theoretical analysis presented below.

To account for the dynamics at the lowest lattice depths, the Gross-Pitaevskii equation
can be safely used because the gas is in the condensate phase. We performed a numerical simulation up to a lattice depth of 7 $E_R$ in order to describe the observed data. The interaction term takes into account short range contact interaction and non local DDI \cite{suppmat}. Concerning DDI we only include the spin conserving terms (see Eq. (\ref{heisenberglike})).

\begin{figure}
\centering
\includegraphics[width= 7 cm]{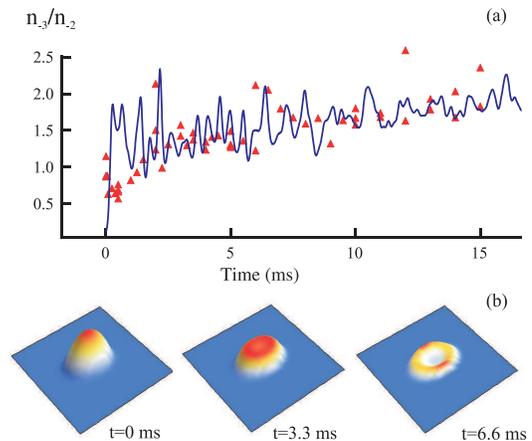}
\caption{Results of numerical simulation using the Gross-Pitaevskii equation. a) Spin dynamics for the deepest lattice depth that could be simulated (7 $E_R$). The (red) triangles are experimental data. b) Spatial analysis of spin dynamics, showing a cut of the density of the atoms in the $m_s=-2$ state  (with no lattice on). }
\label{theory}
\end{figure}

Simulations display a complex behavior, see Fig. \ref{theory} for 7 $E_R$. The general trend of the spin dynamics, showing a slow drift with a $\approx14$ ms characteristic $1/e$ time, is well reproduced by the simulation. By fitting both curves (experimental and numerical) by an exponential, a good agreement for the amplitude is found (see Fig.~\ref{AmpFreq}a)).

In addition, Fourier analysis of the numerical results (up to 15 ms) displays a lowest resolved frequency in good agreement with the largest experimentally observed frequency at a lattice depth of $3~E_R$ and $7~E_R$ (see Fig.~\ref{AmpFreq}b)). Nevertheless, the strong damping observed in the experiment is not reproduced by the zero-temperature simulation. For very low lattice depths (below $3E_R$), simulations show that both contact and dipole interactions contribute to the spin-dynamics. In particular if we set DDI to zero  we
numerically observe a spin dynamics frequency which is roughly twice faster.
This illustrates the interplay between dipolar and contact interactions in the superfluid regime.

We also investigate numerically the spatial dependence of the dynamics
in Fig.~\ref{theory}b), which shows a cut of the density of the atoms in the $m_s=-2$ Zeeman state in absence of an optical lattice, for three evolution times. Similar results are obtained at small lattice depths. The dynamics is inhomogeneous, due to a nontrivial interplay between contact and dipolar interactions. While spin-exchange interactions due to contact interactions are larger in the center of the cloud, due to higher density, dipolar interactions are stronger in the outskirts \cite{Giovanazzi}. The observed dynamics is faster in the center, which illustrates the dominant role of contact interactions at low lattice depths.

We now turn to our theoretical analysis at large lattice depth, where the system is not superfluid and one should go beyond the Gross-Pitaevskii mean-field description. We focus on the frequencies of the two oscillations observed in the Mott phase. We stress that for the largest lattice depth, the experiment is performed in a regime where tunneling is practically absent, and super-exchange interactions are exponentially reduced, as shown in Fig.\ref{AmpFreq}b).

We interpret the upper frequency as a result of intra-site spin-exchange dynamics ($|-2,-2\rangle \rightarrow \frac{1}{\sqrt{2}}(|-3,-1\rangle + |-1,-3\rangle$) arising from spin-dependent contact interaction in doubly-occupied sites. The observed frequency is in good agreement with the theoretical frequency $\frac{4 \pi \hbar^2}{2 \sqrt{2} m}(a_6-a_4)n_0$ (see black solid line in Fig.\ref{AmpFreq}b)). Here, $n_0$ is the peak density in a doubly-occupied lattice site, $m$ the mass of the atoms; $a_6$ and $a_4$ the scattering length of $S=6$ and $S=4$ molecular channels  respectively \cite{feshbachchrome, pasquiou2010}. As also shown in Fig.\ref{AmpFreq}b), weaker on-site confinement in shallower lattices reduces the density $n_0$, and therefore the observed frequency. 

As for the lower frequency, we associate it to non-local DDI between doublons. Indeed, as  reported in our previous work \cite{3DSpinExchange}, this oscillation disappears in presence of a strong magnetic field gradient, which proves that it originates from inter-site interactions; in addition, the oscillation is also absent when doubly occupied sites are emptied \cite{suppmat}. 

We will now describe a perturbative model to account for  the many-body interactions between doublons due to intersite DDI.
We first assume that, after the damping of the fast oscillations, the ensemble of doublons is in
a statistical mixture of $\left|S=6,m_s=4\right\rangle$ and $\left|S=4,m_s=4\right\rangle$. We then describe the spin-dynamics due to inter-site DDI between doubly occupied sites by the following model, inspired from \cite{abragam,porto2015}. We calculate the time evolution of the population $N_{-2}$ in the
state $m_s=-2$ using perturbation theory in the Heisenberg picture. The many-body Hamiltonian takes into account the interaction
of one doubly occupied lattice-site $i$ with all its neighbors $j$ by pair-wise DDI. Using Taylor expansion, the formal expression of the population reads $N_{-2}(t)=\sum_{n=0}^{+\infty}M_{n}t^n$. While $M_1$  and $M_3$ vanish, the second moment $M_2$ describing spin dynamics up to the second order of perturbation reads:

\begin{equation}
M_2= - \sum_{j \neq i} V_{dd}(r_{i,j})^2 /\hbar^2
\label{ratestrong}
\end{equation}
where $V_{dd}(r_{i,j})$ is the dipolar spin-exchange matrix element between sites $i$ and $j$ \cite{suppmat}.

From $M_2$, we first extract an estimate of the spin oscillation frequency  $\nu$:  $\cos^2(\pi \nu t)\approx 1 -M_2 t^2$. However, as shown in \cite{abragam}, the higher order terms are expected to lead to a reduction of
spin dynamics amplitude and a decrease of the quasi period by a factor of typically two. 
We indeed recover these features by taking into account the moment up to the fourth order $M_4$ (following the calculation in \cite{abragam}). We apply this perturbative approach up to fourth order to the case of an assembly of doublons in state $\left|S=6,m_s=4\right\rangle$, for which intersite spin-exchange interactions are strongest. 

The result is displayed in  Fig \ref{AmpFreq}b), and shows good agreement with the observed frequencies, which illustrates the relevance of our model in the deep Mott regime. The observed reduction of the frequency at lower lattice depth may be the consequence of a reduction of the doubly-occupied sites Mott plateau, leading to either stronger border effects, or more defects. The reduction of the frequency in the presence of holes is also an expected feature \cite{abragam}. The damping (also observed in \cite{abragam}) is not reproduced by our theoretical model, which can only
account for dynamics at short times. In general, it should be stressed that the many-body system which we study
here is extremely challenging for non-perturbative many-body simulations,
due to the large spin and the immense Hilbert space which has to be taken into account.

In this paper we have explored spin dynamics of chromium atoms loaded in an optical lattice as a function of the lattice depth. Superfluid and Mott regimes lead to markedly different features. Experimentally, the spin dynamics evolves smoothly when we vary the lattice depth.  However, our analysis shows that the impact of DDI on spin dynamics is drastically different in the two regimes: whereas in the weakly-interacting regime, dipolar interactions are described by a mean-field associated to a geometrical average, in the strongly correlated regime, dipole-dipole couplings from one site to the other sites add quadratically. One important consequence is that spin dynamics mediated by dipolar interactions should be a border effect in the mean-field regime, but a bulk effect in the strongly correlated gas \cite{suppmat}. Space-resolved measurements could therefore be a very interesting way to discriminate both regimes.

Whereas the two extreme regimes of shallow and deep lattice depths are qualitatively well described by our theoretical models, one of the novelties of the paper is the experimental study of the regime close to the Mott to superfluid transition. In this case, exchange processes due to dipolar interactions, spin-exchange due to contact interactions, and super-exchange interactions, may all contribute to the dynamics on the same footing. Our experiment provides first insights into the coupled out-of-equilibium magnetic and transport properties of such a strongly correlated gas, which challenges theoretical descriptions. Our study thus pioneers the study of the magnetic properties of an exotic novel quantum many-body system made of large spin dipolar particles.

Acknowledgements: This work was supported by Conseil R\'{e}gional d'Ile-de-France under DIM Nano-K / IFRAF, Minist\`{e}re de l'Enseignement Sup\'{e}rieur et de la Recherche within CPER Contract, and by Universit\'e Sorbonne Paris Cit\'e (USPC). P. P. acknowledges funding by the Indo-French Centre for the Promotion of Advanced Research - CEFIPRA. We thank Martin Robert-de-Saint-Vincent for his critical reading of the manuscript.

\newpage
\null
\centerline{\Large \textbf{Supplemental material}}

\vspace{0.5cm}

\subsection{GPE Simulation}

To account for the dynamics at the lowest lattice depths, we use the Gross-Pitaevskii equation. We performed a numerical simulation in order to interpret the observed data.
The Gross-Pitaevskii equation for this system reads:
\begin{eqnarray}
i\hbar\frac{\partial}{\partial t}\psi_n(t,\vec r)=\Big(H^0_{n,l}+ \nonumber \\
+ \int \rho_{q,k}(t,\vec r \,') V_{ n,q,k,l}(\vec r - \vec r \,')d^3r'\Big)\psi_l(t,\vec r)
\label{GPeq}
\end{eqnarray}
where $\psi_n(t,\vec r)$ is the wave function for the $n$th spin component in the $z$ direction, $\rho_{q,k}(t,r)=
\psi^*_q(t, \vec r)\psi_k(t, \vec r)$ is the one-body density matrix and $V_{ n,q,k,l}(\vec r)$ is the interaction term.
The summation on repeated indices is implied.

The term $H ^0_{n,l}$ is the single particle term of the Hamiltonian and reads
\begin{equation}
H^0_{n,l} = \delta_{n,l}\left(-\frac{\hbar^2}{2m}\Delta+ V_{\rm ext}(\vec r)\right)+ g\mu_B B S^z_{n,l}\
\end{equation}
This term includes the kinetic energy and potential energy which contains the external trapping and
the magnetic potentials. In order to take into account the lattice potential we renormalize the mass \cite{MassRenormalization}.

The interaction term takes into account short range contact interaction and
DDI. The short range one is written as
$V_{\rm sr}=\sum_{j=0,2,4,6}\bar g_jP_j$, where $P_j=\sum_{n=-j}^j|j,n\rangle\langle j,n|$ is the projector on the molecular state of total spin $j$. To account for the lattice, we use $\bar g_j=\frac{g_j}{(2\pi)^{3/2}{\bar \sigma}^3}$, where $g_J=\frac{4 \pi \hbar^2}{m} a_J$ ($a_J$ being the scattering length in channel $J$), and $\bar \sigma$  is the geometrical average width of the lattice site.

Concerning DDI we only
include the spin conserving terms since in presence of a lattice the
dipolar relaxation is strongly inhibited \cite{ResonancesSM} (see Eq.(\ref{heisenberglike})). Dipolar interactions have a non-local character
which is explicitly taken into account in Eq. (\ref{GPeq}). Due to geometrical averaging, dipolar interactions are typically smaller in the center of the cloud compared to the outskirts. Dipolar coupling constant
does not need to be renormalized.

\subsection{Super-exchange}

The super-exchange mechanisms in the Mott regime for a spinor gas become rather involved due to the multi-channel character of the interaction. A large number of additional spin-exchange phenomena can thus arise in addition to the exchange of the spins found in spin 1/2 systems.
For example, for a spin 3, a two-body state $|2,1\rangle$ (where one site contains an atom in state with $m_s=2$, and the next site contains an atom with $m_s=1$) could become $|1,2\rangle$ or $|3,0\rangle$ or $|0,3\rangle$.
The matrix elements describing tunneling-assisted super-exchange with spin-dependent contact interactions read:
\begin{eqnarray}
U_{m_1,m_2}^{m'_1,m'_2}= 
 \sum_{S,m_S}- \frac{J^2}{U_S} \langle m_1,m_2|S,m_S \rangle \langle S,m_S|m'_1,m'_2 \rangle
\end{eqnarray}
where $|S,m_S \rangle $ is the molecular state of total spin $S$ with projection $m_S$,  and
$\langle S,m_S|m'_1,m'_2\rangle$ are Clebsch-Gordan coefficients. $m_{i=(1,2)}$ represent atomic magnetic states. $J$ is the ($m_i$ independent) tunneling matrix element between nearest sites, and $U_S$ the spin-dependent on-site contact energy (which only depends on S).

\begin{figure}
\centering
\includegraphics[width= 7 cm]{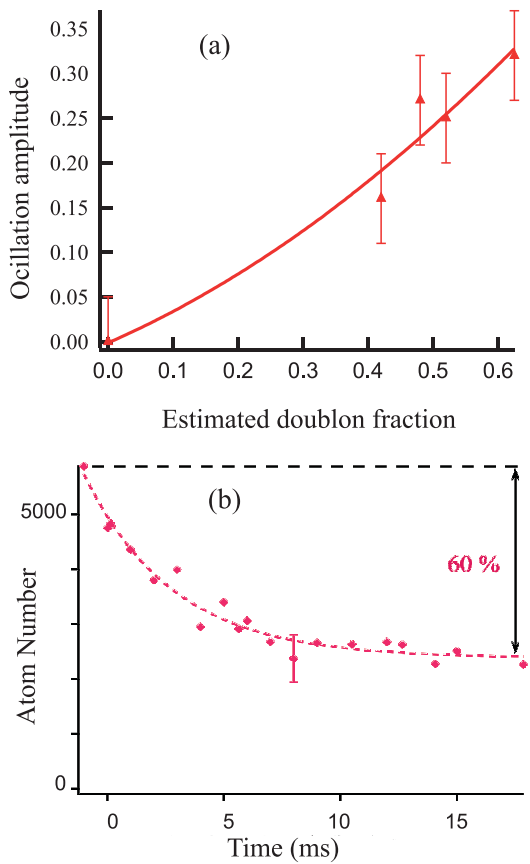}
\caption{(a) Oscillation amplitude as a function of the fraction of doublons in the cloud. The data point at the origin corresponds to a situation where all doublons are removed using dipolar relaxation at large magnetic field. The other points correspond to different temperatures. (b) For each point, the doublon fraction is independently estimated by measuring the fraction of atoms which disappear when doubly occupied sites are emptied using dipolar relaxation at large magnetic fields}
\label{doublondependence}
\end{figure}

\subsection{Spin-exchange as a function of temperature}

To demonstrate that the slow spin oscillations which are observed in the experiment at large lattice depth are due to interactions between doubly-occupied sites, we have varied the number of doublons by modifying the temperature of the gas before it is loaded in the optical lattice. This is achieved by tuning the value of the trap depth at the end of forced evaporation. While the temperature is always kept below the energy gap for band excitation in the optical lattice (corresponding to typically 2.5 $\mu$K), the initial temperature has a strong impact on the atom number distribution in the lattice sites. 

For each given temperature, we characterize the lattice atom distribution by measuring the fraction of atoms located in doubly-occupied sites. To this end, we use the following protocol. The magnetic field is set to a value $> 1$G such that the Zeeman energy significantly exceeds the trap depth. Atoms are promoted to the Zeeman excited state $m_s=3$ using a 5 ms radio-frequency sweep. At such a large magnetic field, dipolar relaxation between a pair of atoms is a local process \cite{pasquiou2010SM}, which releases an energy that is large enough for both atoms to leave the trap. Then, as shown in Fig. \ref{doublondependence} (b), the number of atoms in the lattice rapidly decreases, towards a steady-state non-zero value. This is due to dipolar relaxation occuring in doubly occupied sites, while isolated atoms are immune from dipolar relaxation. The measurement of the fractional loss after dipolar relaxation is therefore a measurement of the fraction of atoms located in doubly occupied sites.

For each temperature, we have also performed a spin-dynamics experiment identical to the one described in the main part of the paper (i.e. after atoms are promoted to state $ms=-2$). We then observe a similar slow spin oscillation, whose amplitude is plotted in Fig. \ref{doublondependence}a) as a function of the estimated number of doublons. This figure demonstrates that the oscillation which is seen in the experiment has an amplitude which increases with the number of doublons. As also reported in \cite{3DSpinExchangeSM}, we observe no oscillation when the doubly occupied sites are emptied using dipolar relaxation at large magnetic field before studying spin dynamics from state $m_s=-2$. Taken together, these measurements therefore confirm that the slow oscillations are due to the interactions between doubly occupied sites (the non-local character of the interaction at play is demonstrated in \cite{3DSpinExchangeSM}).

We also stress that the measurements shown in Fig \ref{doublondependence} (a) indicate that spin oscillations survive at relatively large temperature, corresponding to clouds above the BEC critical temperature before they are loaded in the lattice. The oscillations which we see in our experiment thus survive despite the presence of the disorder in the number of atoms in different lattice sites which necessarily occurs at non-zero temperature.

\subsection{Spin-exchange in a 2D lattice}

The main part of this paper presents the study of spin-exchange dynamics due to intersite dipolar interactions for a range of lattice depths spanning the Mott to superfluid transition. We find that the situation is particularly original and interesting in the superfluid regime, as spin dynamics is then driven by an Heisenberg-like Hamiltonian despite the fact that the atoms are not in an insulating state. Unfortunately, we find that magnetization-changing collisions due to dipolar relaxation are not completely negligible at the lowest lattice depths (see the inset of Fig.\ref{AmpFreq}a)). This is because the Zeeman energy is then similar to the gap for band excitation, so that dipolar relaxation is energetically allowed \cite{ResonancesSM}. Although the agreement between our experimental data and the mean-field simulation based on the Gross-Pitaevskii equation neglecting dipolar relaxation is still satisfactory, the parallel to the Heisenberg model of magnetism may then be questionable, because the Heisenberg model of magnetism does not include magnetization-changing collisions.  

\begin{figure}
\centering
\includegraphics[width= 7 cm]{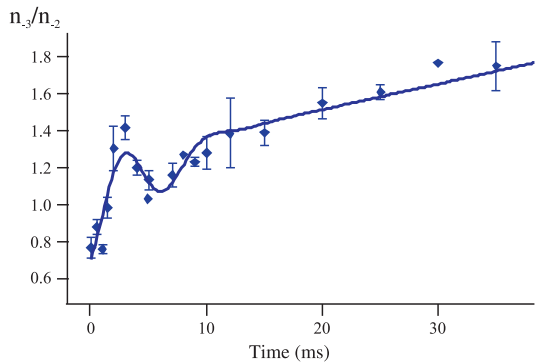}
\caption{Time evolution of observable $n_{-3}/n_{-2}$ for a 2D array of 1D quantum gases produced by loading atoms in a deep 2D optical lattice.}
\label{tubesoscillation}
\end{figure}

To probe spin-dynamics in a mean-field (superfluid) regime, while maintaining a constant magnetization in time, we have
therefore performed additional similar experiments in a deep 2D optical lattice, where dipolar relaxation is strongly
reduced due to a small density of states at the Larmor energy \cite{CrtubeSM}. Then the atoms form a periodic array of one-dimensional quantum gases. As shown in figure \ref{tubesoscillation}, after the atoms are promoted to state $ms=-2$, we observe spin dynamics with an oscillatory behavior, at constant magnetization, similar to what is obtained in a 3D optical lattice. The oscillation frequency which is deduced from a fit using a damped sinusoidal function, 165 Hz, is significantly smaller than the frequency measured in the deep 3D optical lattice, but comparable to the ones measured for lower 3D lattice depths.

The theoretical interpretation of spin-dynamics in this 2D lattice is difficult, because dipolar interaction may include intersite couplings between the 1D quantum gases created by the optical lattice, and interactions within one tube of atoms. In addition, spin-dynamics due to on-site spin-dependent contact interactions may not be negligible. However, the fact that the observations are qualitatively the same above the 3D Mott transition in a 3D lattice and in an array of 1D gases in a 2D lattice is an interesting feature, which shows that the basic out-of-equilibrium magnetic properties of the insulating state are preserved in presence of transport. 

\subsection{Dipolar interactions in the mean-field and in the strongly correlated regimes}

As shown in the main part of the paper, dipolar interactions lead in principle to drastically different behaviors in the mean-field and in the strongly correlated regimes. In the mean-field regime, dipolar interactions between atoms create a non-local mean-field, determined by a global average over space
\begin{equation}
\phi(\vec{r}) \propto \int d^3r' V_{dd}(r-r') n(r')
\label{meanfield}
\end{equation}
To simplify this specific discussion, the tensorial nature of $V_{dd}$ is overlooked; see Eq. (\ref{GPeq}). Eq. (\ref{meanfield}) in practice implies that for a homogeneous system, dipolar interactions vanish and should not contribute to spin dynamics. Therefore,
spin-dynamics due to dipolar interaction is a border effect in the mean-field regime. 

On the other hand, in the Mott regime, according to second order perturbation theory applied to an initial state where each lattice site is populated by one atom in state $m_s=-2$, spin dynamics due to dipolar interactions is qualitatively described by the following effective rate at short times
\begin{equation}
\Gamma= \frac{1}{\hbar} \sqrt{\sum_{i<j} V_{dd}(r_{i,j})^2}
\label{manybody}
\end{equation}
where
\begin{eqnarray}
V_{dd}(r_{i,j})=\frac{d^2 (1-3\cos^2(\theta_{i,j})}{4 r_{i,j}^3}   \nonumber\\
\left\langle 3,-2;3,-2\right|S_i^+ S_j^-\left|3,-1;3,-3\right\rangle 
\end{eqnarray}
 describes spin exchange dipole-dipole interactions between two $m_s=-2$ atoms in sites $i$ and $j$. When we analyze the case of doubly occupied sites, we consider spin-exchange between two $S=6$ pairs of atoms in state $\left|S=6,m_S=-4\right\rangle$, and we use
\begin{eqnarray}
V_{dd}(r_{i,j})=\frac{d^2 (1-3\cos^2(\theta_{i,j})}{4 r_{i,j}^3}  \times \nonumber\\
\left\langle 6,-4;6,-4\right|S_i^+ S_j^-\left|6,-3;6,-5\right\rangle 
\end{eqnarray}

The difference between Eq. \ref{meanfield} and Eq. \ref{manybody} is striking and illustrates the difference between the two regimes. In the superfluid regime, the effect of dipolar interactions results from a geometrical averaging, whereas in the strongly correlated regime, it results from a quadratic averaging. As a consequence, in the strongly correlated regime, spin-exchange due to dipolar interactions does not vanish even for a homogeneous system. 

To better understand the difference between Eq. \ref{meanfield} and Eq. \ref{manybody}, we propose the following physical interpretation. The initial many-body state is $\Psi_0=\left|-2,-2,....,-2\right\rangle$, where the spin state of each atom in each lattice site is explicitly written down: each site contains one $m_s=-2$ atom. Dipolar interaction is a sum of pair-wise interactions (see Eq. \ref{heisenberglike}) between atoms in sites $i$ and $j$. Therefore $\Psi_0$ is directly coupled to the following many-body state:
\begin{eqnarray}
\Psi_1=\frac{1}{\sqrt{\sum_{i<j} V_{dd}(r_{ij})^2}} \times \nonumber\\
\sum_{i<j} V_{dd}(r_{ij}) \left|-2,...,i:-1,..,j:-3...,-2\right\rangle 
\label{statecoupled}
\end{eqnarray}
In the state $\left|-2,...,i:-1,..,j:-3...,-2\right\rangle$ all sites contain a $m_s=-2$ atom, except site $i$ (resp. $j$) which contains a $m_s=-1$ (resp. -3) atom. The rate of coupling between $\Psi_0$ and $\Psi_1$ is given by Eq. \ref{manybody}. This many-body physical picture clearly demonstrates that the initial state is coupled to a state which shows spin entanglement. As a consequence, spin dynamics in this regime cannot be grasped by the Gross-Pitaevskii equation and by Eq. \ref{meanfield}.

This discussion indicates that the difference between the regime at low lattice depth and the situation at large lattice depth could be further experimentally investigated, either by measuring the growth of entanglement (for large lattice depth), or by performing position resolved spin dynamics measurements.

\end{document}